\documentclass[aps,prl,preprint,groupedaddress]{revtex4-1}
\usepackage{graphics}
\usepackage{graphicx}
\begin{document}

%\preprint{}

\title{Electron scale nested quadrupole Hall
field in Cluster observations of magnetic reconnection}


\author{Neeraj Jain}
%\thanks{}
\affiliation{Max Planck Institute for Solar System Research, Justus-von-Liebig-Weg-3, G\"{o}ttingen, Germany.}

\author{A. Surjalal Sharma}
%\thanks{}
\affiliation{Department of Astronomy, University of Maryland, College Park, MD 20742, USA.}

\date{\today}

\begin{abstract}
% insert abstract her
This Letter presents the first evidence of
a new and
unique feature of spontaneous reconnection at multiple sites in electron current
sheet, viz. “nested quadrupole” structure of Hall field at electron scales, in Cluster observations. 
The new nested quadrupole  
is a consequence of
electron scale processes in reconnection.
Whistler response of the upstream plasma to the interaction of electron flows
from neighboring reconnection sites produces a large scale quadrupole Hall
field enclosing the quadrupole fields of the multiple sites,   thus forming a
nested
structure. 
Electron-magnetohydrodynamic simulations of an electron
current sheet yields mechanism of the formation of nested quadrupole.
\end{abstract}

\pacs{}
%\keywords{}

\maketitle

Magnetic reconnection 
is a fundamental process  for the 
fast release of  magnetic energy into  kinetic and thermal energy
in laboratory, space and astrophysical plasmas.
Collisionless reconnection develops in thin current sheets with thicknesses
comparable to the 
electron skin depth
 $d_e(=c/\omega_{pe})$. The electron current sheet (ECS) with thickness $\sim
d_e$ 
is embedded inside an ion current sheet with thickness $\sim
d_i(=c/\omega_{pi})$. The electron and ion dynamics are decoupled at this scale
and the plasma is no longer frozen in the 
magnetic field, thus enabling reconnection. 
The Hall current due to the differential
 flow of ions and electrons in the reconnection region generates an out-of-plane
magnetic field
with a quadrupolar structure \cite{sonnerup79,mandt94}, which 
will be referred to as the Hall field. The quadrupole structure of the  Hall
field is an essential feature of collisionless reconnection and has been
detected in 
space observations \cite{wygant05,borg05,asano04}, laboratory 
experiments \cite{ren05} and simulations \cite{hesse01a}. 

\begin{figure}
 \includegraphics[width=0.5\textwidth,height=.5\textheight]
 {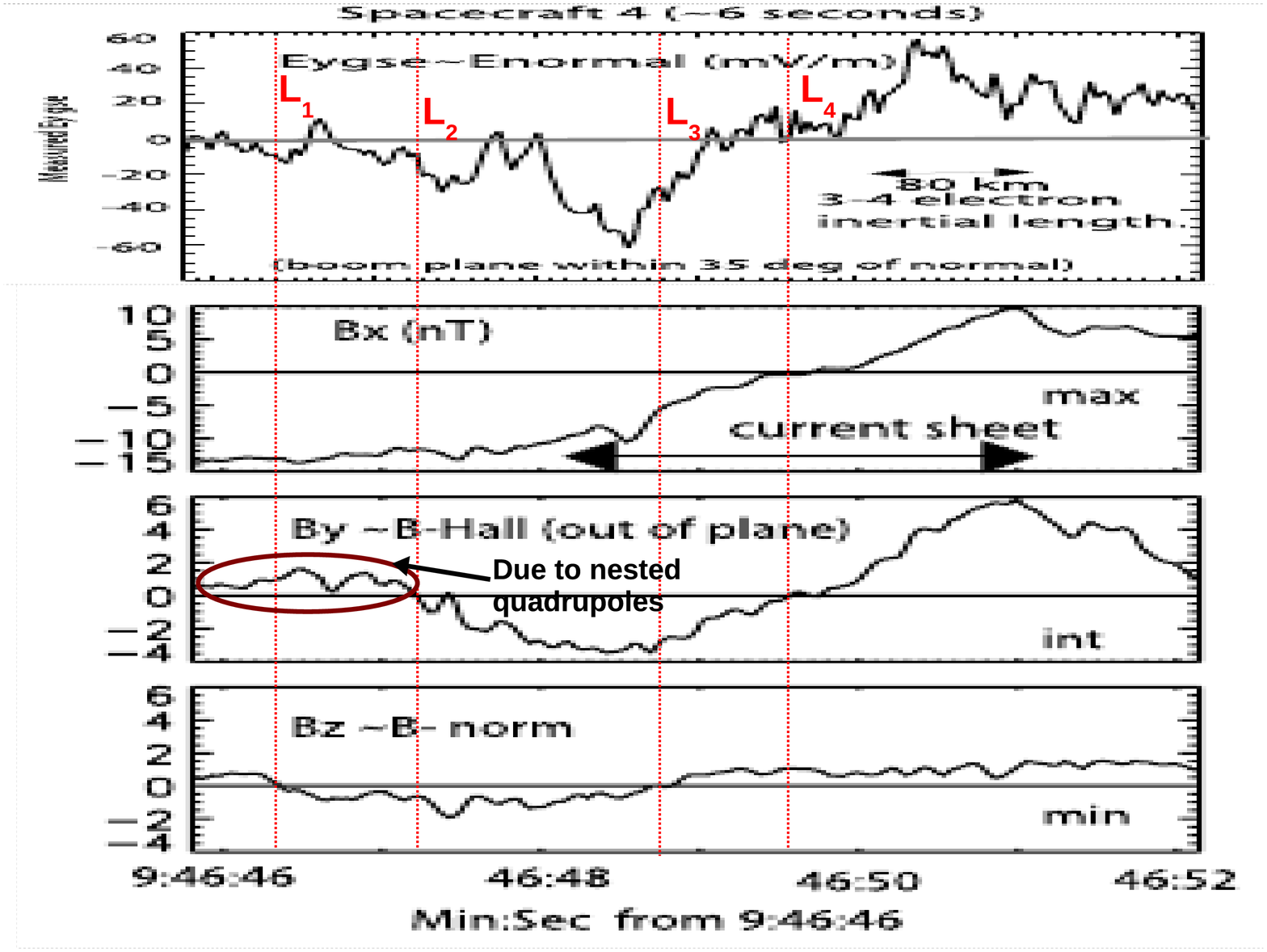}
\caption{Observation of electron scale current sheet by Cluster 
(adopted from Fig. 3a of  
\cite{wygant05}). (top
panel) $E_{yGSE}$ ($y$-component    of electric field in GSE coordinate  system
).  (Three  bottom panels) $x$, $y$ and $z$ components of magnetic field   in
boundary normal coordinate system. Vertical dashed lines ($L_1-L_4$)
mark the  zero crossings of the 
magnetic field components.
\label{wygantmatch1}}
\end{figure}

The electron current sheet  is susceptible to secondary tearing
instabilities
which lead to spontaneous reconnection at multiple sites in ECS
\cite{daughton06}. The interaction of neighboring sites leads to a new and
unique
feature, viz. nested quadrupole structure of  
the Hall field \cite{jain09},  
unlike the single quadrupole 
in the case of  reconnection at a single site.  This feature 
arises in 
electron current sheets with a  
thickness ($\sim$ few $ d_e$) which is small compared to its  
extent  ($\sim $ few $d_i$). 
Such current sheets are unstable to tearing instability, with a growth rate that
has a maximum when the perturbation has scale length of a few $d_e$
\cite{jain09,attico00}, thus leading to reconnection at multiple sites.
This Letter presents the first evidence of a
nested quadrupole structure of Hall field 
 in the Cluster observations
of an electron scale current sheet in Earth's magnetotail \cite{wygant05}. The
Cluster spacecraft crossed the reconnection region at distances of $\sim18R_E$
in Earth's magnetotail on 1 October 2001. 
Among the four spacecraft SC4 was closest to the X-line and crossed the current
sheet on the earthward side between 09:46:48 and 09:46:51 UT, and the 
profiles of 
electric and magnetic field
are shown in Fig. \ref{wygantmatch1} (Fig. 3 in Ref. \citep{wygant05}).
The change in sign of the magnetic field components are critical to the structure of the Hall field and the time marks for these are shown by the vertical dashed lines in Fig. 1, viz. $L_1$ for $B_z$, $L_2$ for $B_y$, $L_3$ for  $B_z$, and $L_4$ for $B_x$ and $B_y$.

\begin{figure}
\includegraphics[width=0.5\textwidth,height=0.3\textheight]{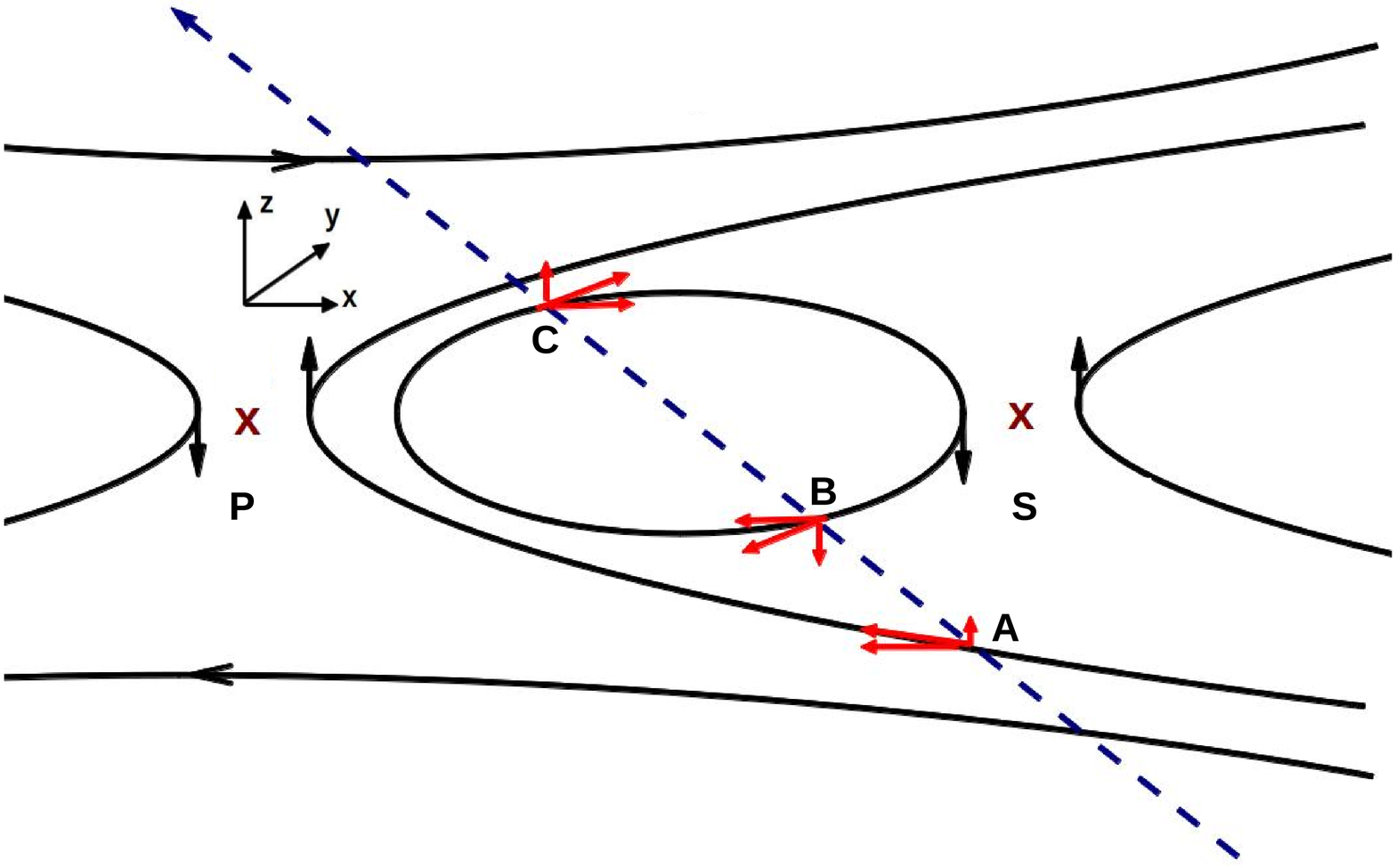}
\caption{A schematic of reconnection  at a primary (P) and a
secondary
(S) sites. Components of the magnetic
field in $x$ and $z$ directions are shown by red arrows at select locations (A, B and C) on
a spacecraft trajectory (blue dashed line).   
\label{fig:schematic}}
\end{figure}

A schematic of the magnetic field structure corresponding to the Cluster
observations (Fig. 1)
is shown in Fig. \ref{fig:schematic}, and consists
of a primary site, with X-point at P, and a secondary site with X-point at S.
In the standard picture of 2-D reconnection with a single reconnection site, i. e., in the absence of the secondary sites, 
$B_z$ should have the same sign on any one side (tail-ward or earthward) of the $y-z$ plane containing the X-point P, and change sign only
when spacecraft crosses this plane.  
But for such a passage  by a spacecraft, the peak of the out-of-plane Hall  
field ($B_y$) should not
coincide with the
zero crossing of the normal magnetic field ($B_z$). This is because the peaks of
the Hall field are located away from this plane, as 
seen in the Cluster data
(Line
$L_3$). Thus the change of sign of $B_z$ at $L_3$ coinciding with the peak of the Hall field is not consistent with reconnection at a single site.
The change of sign of $B_z$ without crossing an X-point is possible,
however, when spacecraft crosses the current sheet between a primary (P) and a
secondary (S) reconnection sites,
e.g., along the dashed line in Fig. \ref{fig:schematic}. The spacecraft first
encounters  magnetic field lines (at A with positive $B_z$) reconnected at the
primary site P but not reconnected at the secondary site S.  It then encounters
the field lines of the magnetic island formed due to the reconnection both at P
and S, first in the the region below the plane containing the primary and secondary sites, viz. the south lobe (at B with negative $B_z$) and then in the north lobe
(at C with positive $B_z$). Although the presence of a magnetic island between the two
reconnection sites P and S is enough for the sign reversal of $B_z$ at $L_3$ and
$B_y$ at $L_4$, the small positive $B_z$ on the left of $L_1$ and $B_y$ on the
left of $L_2$ additionally require the weak or secondary site to be inside the region created by the  dominant or primary site.     

\begin{figure}
\includegraphics[width=0.5\textwidth,height=0.4\textheight]{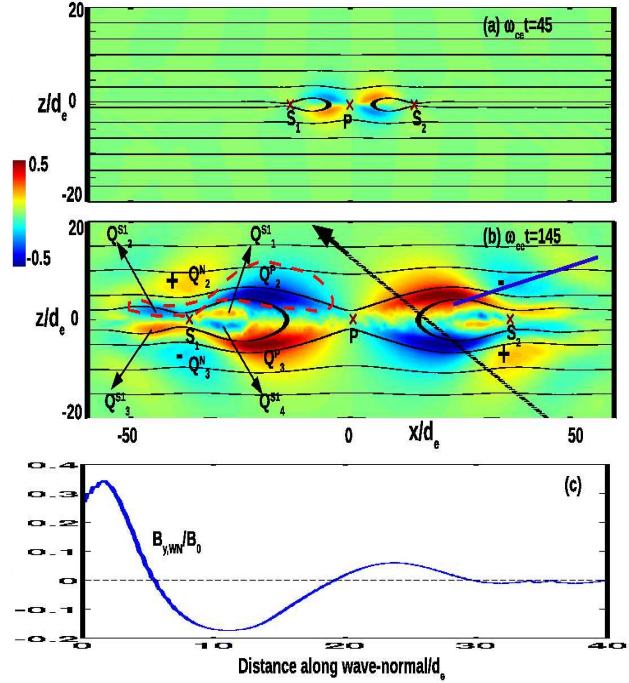}
\caption{Magnetic field lines (black) plotted over color coded 
$B_y$ at $\omega_{ce}t=45$ (a) and 
=145 (b). 
The primary (P) and secondary 
(S$_1$ and S$_2$) sites are marked by cross ($\times$). At 
$\omega_{ce}t=145$, outward oblique propagation of whistlers from secondary sites
forms a new quadrupole, 
marked with '+' and '-'.  In (b), the poles of primary, secondary and new
quadrupoles are marked by Q's (see text for definition) in the left half.   
The red-dashed loop
encloses a negative pole of the extended quadrupole. The blue line in
the top-right quadrant is at an angle of 19.5
degrees with the background magnetic field along $+x$ and approximates the wave
normal.  The profile of $B_y$ along the wave normal is shown in (c). The black line with an arrow in (b) shows a possible trajectory of Cluster
spacecraft.    
\label{fig:bz_evolution}}
\end{figure} 

\begin{figure}
 \includegraphics[width=0.5\textwidth,height=.5\textheight]
 {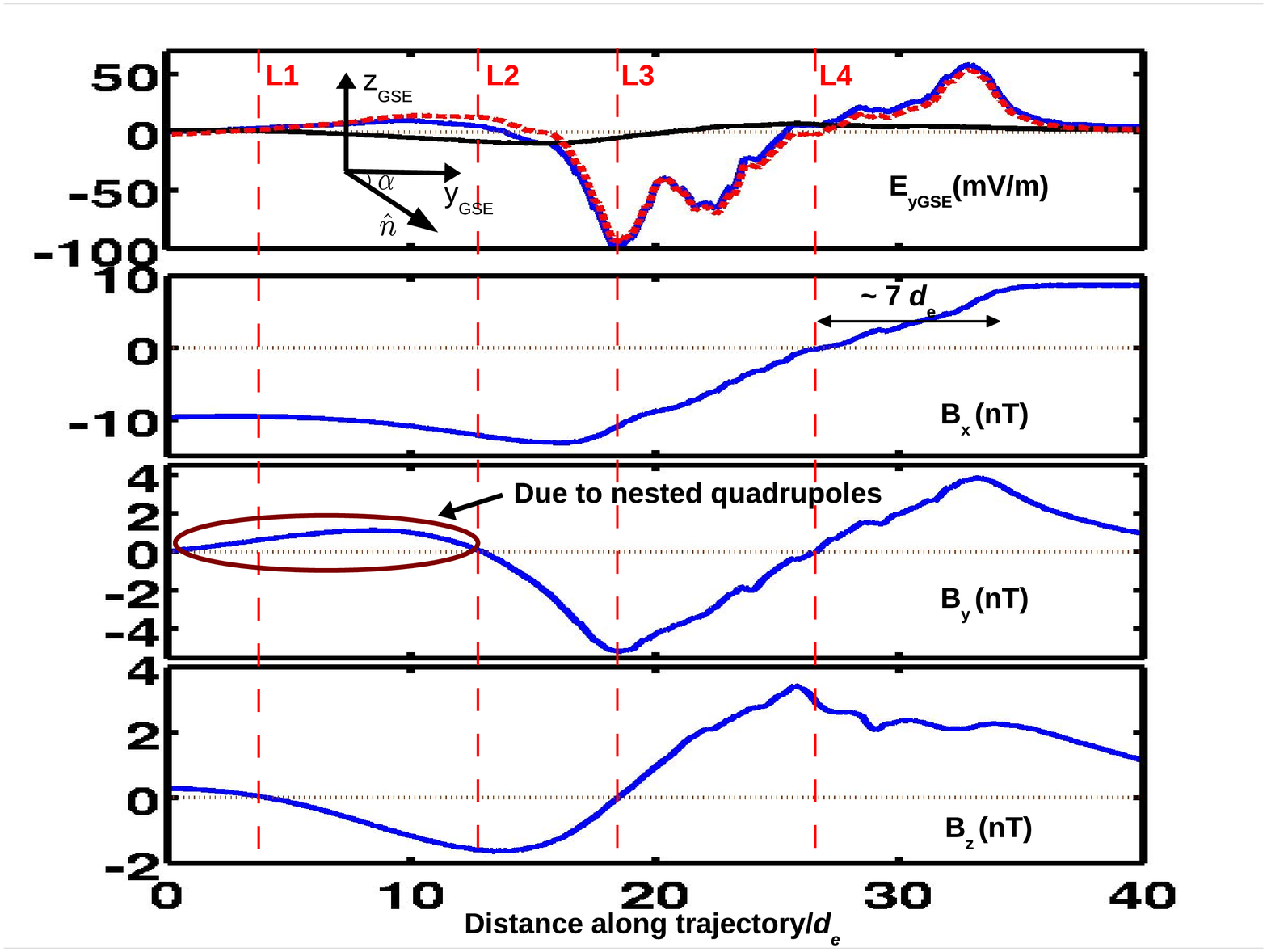}
\caption{Simulated electric and magnetic field profiles along the trajectory  shown in Fig.
\ref{fig:bz_evolution}b. (top panel) The $y$-component    of electric
field ($E_{yGSE}$, blue) transformed from the simulation or boundary normal to GSE
coordinate system, the normal component
(red) and current aligned (black) electric field of $E_{yGSE}$. Also shown is
the boundary normal vector in the $y_{GSE}$-$z_{GSE}$ plane. (Three bottom
panels) The $x$, $y$ and $z$ components of magnetic field in
boundary normal coordinate system. Vertical dashed lines ($L_1-L_4$)
mark the  zero crossings of the 
magnetic field components. 
\label{wygantmatch2}}
\end{figure}

The correspondence of the Cluster observations of the magnetic field (Fig.
\ref{wygantmatch1}) 
to reconnection at multiple sites is modeled using  
electron-magnetohydrodynamic (EMHD) simulations of spontaneous reconnection in
an ECS developing into primary and secondary sites \cite{jain09},
shown  in     
Fig. \ref{fig:bz_evolution}. Here length is normalized by $d_e$, magnetic field
by the asymptotic value $B_0$ and time by $\omega_{ce}^{-1}=(eB_0/m_e)^{-1}$. 
The profiles of electric and magnetic fields in the simulations along a possible
 spacecraft trajectory (shown by the thick line with an arrow in   Fig.
\ref{fig:bz_evolution}b) are shown in Fig. \ref{wygantmatch2}. 
Fig. \ref{fig:bz_evolution}a 
shows the structure of the  
Hall field $B_y$
in the 
early stage ($\omega_{ce}t=45$), which evolves into   the late stage ($\omega_{ce}t=145$),  shown in Fig. \ref{fig:bz_evolution}b. 
At $\omega_{ce}t=45$ the reconnection is dominant at the
primary site (P) in the center
of 
the simulation domain ($x=z=0$). The field lines reconnected at the primary site
P reconnect  again at the secondary sites, (S$_1$ at 
$x\approx-16 d_e$ and S$_2$ at $x\approx 16 d_e$), giving rise
to reconnection at 
multiple sites.
 The quadrupole structure of the out-of-plane magnetic
field is clearly developed around the primary site, while it is not recognizable yet  at
the secondary sites.  At $\omega_{ce}t=145$, the central site remains dominant and
the secondary sites are pushed away by the outflows from the central site.
We label the quadrupole Hall fields associated with
S$_1$, S$_2$ and P by Q$^{S_1}$, Q$^{S_2}$ and Q$^P$, respectively. The new quadrupole, marked as $Q^N$, forms due to the interactions of the inflow to the secondary sites and the outflow from the primary site \cite{jain09}. 
The poles of a quadrupole are numbered counter clockwise beginning with 1 
for the top
right pole to 4 
for the bottom right pole. An
individual pole of a quadrupole is represented by a
subscript to Q's. For example, Q$^{S_1}_1$ represents the top right pole  of the
quadrupole associated with the reconnection site S$_1$.  The poles of the
primary, secondary and  new quadrupoles are marked only in the left half of Fig.
\ref{fig:bz_evolution}b. 
The poles Q$^{S_1}_1$ and Q$^{S_1}_4$ of the secondary quadrupole at S$_1$  penetrates between the poles Q$^{P}_2$ and
Q$^{P}_3$ of primary quadrupole. At the same time, the poles Q$^{S_1}_2$ and
Q$^{S_1}_3$  of the secondary quadrupole at S$_1$ connect to the poles Q$^{P}_2$
and
Q$^{P}_3$ of the primary quadrupole, respectively, thus increasing the extent of the primary quadrupole
Q$^P$. One of the negative pole (Q$^P_2$+Q$^{S_1}_2$) of the extended
quadrupole  is enclosed by a closed loop (red dashed line)  in Fig.
\ref{fig:bz_evolution}b. The extended quadrupole is nested inside the 
new quadrupole (Q$^N$), the poles of which are also marked ('+' and
'-') in Fig. \ref{fig:bz_evolution}b.

A striking feature of spontaneous reconnection at multiple sites is the new
quadrupole
which, unlike the other three quadrupoles in Fig. \ref{fig:bz_evolution}b, is not
directly associated with a reconnection site but arises from  their
interaction.
The physics of the new quadrupole is the whistler response of the upstream
plasma to the interaction of 
inflow to the secondary (weak) sites and outflow from the  primary (dominant) site  \cite{jain09}. 
Because of the magnetic field structure of reconnection, the whistler perturbations are anchored in phase at their origin and propagate away from the reconnection region. The direction of propagation is very well approximated by the wave normal (shown by blue line in Fig. \ref{fig:bz_evolution}b) which is at Storey 
angle of $19.5^{\circ}$ \cite{storey1953,singh2011} with the background magnetic field along $x$. 
Fig. \ref{fig:bz_evolution}c shows the out-of-plane magnetic field $B_{y,WN}$
along the wave normal.  The wave propagates away from the reconnection region while its
amplitude diminishes. The distance between positive and negative peaks is
$\approx
12 d_e$ giving a wavenumber $k d_e\approx 0.25$, as expected for frequency
$\omega=0.1\omega_{ce}$ \cite{singh2007}. 
The extension of the primary 
quadrupole along $x$, and, the formation of a new quadrupole due
to
the whistler perturbation at secondary sites in the manner described above make the overall structure a 
nested  structure of  quadrupoles.

Fig. \ref{wygantmatch2} shows the profiles of electric and magnetic fields 
(in un-normalized units using $B_0=10$ nT and $d_e=20$ km for Cluster
observations) along the trajectory shown in Fig. \ref{fig:bz_evolution}b, as
functions of distance along the trajectory. Similar to Fig. \ref{wygantmatch1}, the vertical dashed lines in this
figure mark the zero crossing of $B_x$ at $L_4$, $B_y$ at $L_2$ and $L_4$, and
$B_z$ at $L_1$
and $L_3$.  
The profiles of the $y$-component of electric field and all components of
magnetic field in Fig. \ref{wygantmatch1} are in the Geocentric Solar Ecliptic (GSE) and boundary normal
coordinate systems, respectively. In the boundary normal coordinate system, $z$ is normal to the current sheet surface, $y$ is along the direction of current and $(x,y,z)$  forms a right handed coordinate system. Since the simulations are in boundary normal
coordinate system, the profile of the electric field in Fig. \ref{wygantmatch2}
is obtained by transforming it from boundary normal to the  
GSE system.
The boundary normal vector
$\hat{n}=-0.05\hat{x}_{GSE}+0.80\hat{y}_{GSE}-0.59\hat{z}_{GSE}$ of the highly
tilted current sheet in Cluster observations is almost in the
$y_{GSE}$-$z_{GSE}$ plane and shown in the top panel of Fig. \ref{wygantmatch2}.
Assuming the simulation current sheet to have the same orientation with respect
to the GSE coordinate system, the $y$-component of the electric field 
in the latter can be obtained from $E_{yGSE}=E_y\sin(\alpha)-E_z\cos(\alpha)$, where $\alpha$ is the angle between the normal vector and the $\hat{y}_{GSE}$, with
$\cos(\alpha)=0.8$. 

The electric and magnetic field profiles in the Cluster observation (Fig.
\ref{wygantmatch1}) and EMHD simulation (Fig. \ref{wygantmatch2}) 
are remarkably similar not only in 
magnitude but also in the scale and pattern of variation.  
The current sheet crossing, represented by the change in $B_x$ from $\approx$-10 nT to $\approx$10 nT in observations
(during $\sim 46:48-46:51$, Fig. \ref{wygantmatch1}) and simulations  (Fig.
\ref{wygantmatch2}), provides more details of the reconnection in the magnetotail. The half thickness of
the current sheet in simulations $\approx 7 d_e$ compares well with the
observed values $\sim 3-5 d_e$. The step-like structures of $B_x$ inside the current
sheet are present both in simulations and observations, and indicate a
filamentary structure in the current sheet. 

Associated with the current sheet crossing, $E_{yGSE}$ and Hall field $B_y$ 
have bipolar forms changing their signs from negative to positive. The positive
and negative peaks of the bipolar structures of $E_{yGSE}$ and $B_y$ in
observations and simulations are very similar. Consistent with the observations, Fig.
\ref{wygantmatch2} shows that $E_{yGSE}$, given by $E_y\sin(\alpha)-E_z\cos(\alpha)$, is dominated
by the normal component of the electric field, $E_z\cos(\alpha)$, due to the tilt
of the current sheet with respect to the GSE coordinate system. Line $L_4$ in Fig.
\ref{wygantmatch1} and Fig. \ref{wygantmatch2} show that $E_{yGSE}$ crosses
zero  earlier than both $B_x$ and $B_y$, which cross zero simultaneously. The
normal component of magnetic field $B_z$ remains positive during the current
sheet crossing but 
is negative just before the current sheet crossing (between $L_1$ and $L_3$).
The zero crossing of $B_z$ at $L_3$ coincides with the edge of the current sheet
and negative peaks of $E_{yGSE}$ and $B_y$. Both $B_z$ and $B_y$ 
have positive values before their first zero crossings at $L_1$ and $L_2$,
respectively. In the simulations, the positive $B_y$ on the left of $L_2$ is due to 
the crossing of a positive pole (marked by '+' on positive $x$-side in Fig.
\ref{fig:bz_evolution}b) of the new (outer) quadrupole structure of $B_y$. The positive $B_y$ on the left of $L_2$
in Fig. \ref{wygantmatch1} can be identified with  the new quadrupole and 
 the Cluster 
observation is consistent with a nested quadrupole structure of the out-of-plane
magnetic field.

The formation of the nested structure of quadrupoles of the 
Hall magnetic field requires not only the presence of multiple sites but also the dominance of one
site over the neighboring sites. 
Simulations with three reconnection sites of equal strength (excited by initializing the simulations with a single wavelength perturbation  with three wavelengths fitting in the length of the simulation box along $x$) show that the out-of-plane
magnetic field does not develop nested structure of quadrupoles. Although quadrupole structure of $B_y$ forms at each reconnection site, the inflow
to one site and outflow from the neighboring site do not interact in the manner 
that results into the nested quadrupole structure.   
In natural situations, e.g., in the magneto-tail,  reconnection at multiple sites is expected, with the one initiated first being dominant over the adjacent sites. Further, in the magnetotail the monotonic decrease of the magnetic field away from Earth (along $x$) will introduce asymmetry among the multiple reconnection sites, thus leading to the nested Hall field.

In the Cluster observations, the total time of crossing $\approx$ 6 sec.,  
close to the ion cyclotron period and thus captured the electron dominated physics of reconnection.
Since these electron scale observations are by a single
spacecraft when the other three spacecrafts were separated by distances much larger
than typical electron scales ($\sim 20$ km), 
the spatial and time variations are not uniquely distinguished.
 However, the EMHD simulations show that the electron scale structures form very
quickly, in a time of the order of tens of electron cyclotron periods,  but
evolve very slowly after their formation \cite{jain09}. Thus the structures
observed by Cluster are consistent with spatial variations as described above. 
The forthcoming multi-spacecraft 
NASA/MMS mission, designed  to resolve the electron scales in
the magnetosphere and to distinguish between spatial and time variations, will provide key details of the spatio-temporal structure.

The nested quadrupole structure of Hall magnetic field identified in Cluster
observations and 
the underlying mechanism revealed by EMHD simulations  
focus only on electron scale processes. 
Many details of the electron scale
physics and the connection to the larger scale ion processes remain unexplored. 
Such studies will require 
new studies of electron scale physics in simulations, experiments and satellite
observations of magnetic reconnection. In particular, the results
presented in this Letter provide a critical step for a deeper understanding of
reconnection at electron scales using new kinetic simulations that resolve the 
electron scales clearly and the 
data for electron scale physics 
from the upcoming NASA/MMS mission.


\begin{thebibliography}{10}%
\makeatletter
\providecommand \@ifxundefined [1]{%
 \ifx #1\undefined \expandafter \@firstoftwo
 \else \expandafter \@secondoftwo
\fi
}%
\providecommand \@ifnum [1]{%
 \ifnum #1\expandafter \@firstoftwo
 \else \expandafter \@secondoftwo
\fi
}%
\providecommand \enquote [1]{``#1''}%
\providecommand \bibnamefont  [1]{#1}%
\providecommand \bibfnamefont [1]{#1}%
\providecommand \citenamefont [1]{#1}%
\providecommand\href[0]{\@sanitize\@href}%
\providecommand\@href[1]{\endgroup\@@startlink{#1}\endgroup\@@href}%
\providecommand\@@href[1]{#1\@@endlink}%
\providecommand \@sanitize [0]{\begingroup\catcode`\&12\catcode`\#12\relax}%
\@ifxundefined \pdfoutput {\@firstoftwo}{%
 \@ifnum{\z@=\pdfoutput}{\@firstoftwo}{\@secondoftwo}%
}{%
 \providecommand\@@startlink[1]{\leavevmode}%
 \providecommand\@@endlink[0]{}%
}{%
 \providecommand\@@startlink[1]{%
  \leavevmode
  \pdfstartlink
   attr{/Border[0 0 1 ]/H/I/C[0 1 1]}%
   user{/Subtype/Link/A<</Type/Action/S/URI/URI(#1)>>}%
  \relax
 }%
 \providecommand\@@endlink[0]{\pdfendlink}%
}%
\providecommand \url  [0]{\begingroup\@sanitize \@url }%
\providecommand \@url [1]{\endgroup\@href {#1}{\urlprefix}}%
\providecommand \urlprefix [0]{URL }%
\providecommand \Eprint[0]{\href }%
\@ifxundefined \urlstyle {%
  \providecommand \doi [1]{doi:\discretionary{}{}{}#1}%
}{%
  \providecommand \doi [0]{doi:\discretionary{}{}{}\begingroup
  \urlstyle{rm}\Url }%
}%
\providecommand \doibase [0]{http://dx.doi.org/}%
\providecommand \Doi[1]{\href{\doibase#1}}%
\providecommand \bibAnnote [3]{%
  \BibitemShut{#1}%
  \begin{quotation}\noindent
    \textsc{Key:}\ #2\\\textsc{Annotation:}\ #3%
  \end{quotation}%
}%
\providecommand \bibAnnoteFile [2]{%
  \IfFileExists{#2}{\bibAnnote {#1} {#2} {\input{#2}}}{}%
}%
\providecommand \typeout [0]{\immediate \write \m@ne }%
\providecommand \selectlanguage [0]{\@gobble}%
\providecommand \bibinfo [0]{\@secondoftwo}%
\providecommand \bibfield [0]{\@secondoftwo}%
\providecommand \translation [1]{[#1]}%
\providecommand \BibitemOpen[0]{}%
\providecommand \bibitemStop [0]{}%
\providecommand \bibitemNoStop [0]{.\EOS\space}%
\providecommand \EOS [0]{\spacefactor3000\relax}%
\providecommand \BibitemShut [1]{\csname bibitem#1\endcsname}%
%</preamble>
\bibitem{sonnerup79}%
  \BibitemOpen
  \bibfield{author}{%
  \bibinfo {author} {\bibfnamefont{B.~U.~O.}\ \bibnamefont{Sonnerup}},\ }%
  \emph{\bibinfo {title} {{in} Solar System Plasma Physics, {vol. 3}}}\
  (\bibinfo {publisher} {North-Holland},\ \bibinfo {address} {Amsterdam},\
  \bibinfo {year} {1979})\ pp.\ \bibinfo {pages} {47--108}%
  \bibAnnoteFile{NoStop}{sonnerup79}%
\bibitem{mandt94}%
  \BibitemOpen
  \bibfield{author}{%
  \bibinfo {author} {\bibfnamefont{M.~E.}\ \bibnamefont{Mandt}}, \bibinfo
  {author} {\bibfnamefont{R.~E.}\ \bibnamefont{Denton}},\ and\ \bibinfo
  {author} {\bibfnamefont{J.~F.}\ \bibnamefont{Drake}},\ }%
  \bibfield{journal}{%
  \bibinfo {journal} {Geophys.\ Res.\ Lett.}\ }%
  \textbf{\bibinfo {volume} {21}},\ \bibinfo {pages} {73} (\bibinfo {year}
  {1994})%
  \bibAnnoteFile{NoStop}{mandt94}%
\bibitem{wygant05}%
  \BibitemOpen
  \bibfield{author}{%
  \bibinfo {author} {\bibfnamefont{J.~R.}\ \bibnamefont{Wygant}}, \bibinfo
  {author} {\bibfnamefont{C.~A.}\ \bibnamefont{Cattell}}, \bibinfo {author}
  {\bibfnamefont{R.}~\bibnamefont{Lysak}}, \bibinfo {author}
  {\bibfnamefont{Y.}~\bibnamefont{Song}}, \bibinfo {author}
  {\bibfnamefont{J.}~\bibnamefont{Dombeck}}, \bibinfo {author}
  {\bibfnamefont{J.}~\bibnamefont{McFadden}}, \bibinfo {author}
  {\bibfnamefont{F.~S.}\ \bibnamefont{Mozer}}, \bibinfo {author}
  {\bibfnamefont{C.~W.}\ \bibnamefont{Carlson}}, \bibinfo {author}
  {\bibfnamefont{G.}~\bibnamefont{Parks}}, \bibinfo {author}
  {\bibfnamefont{E.~A.}\ \bibnamefont{Lucek}}, \bibinfo {author}
  {\bibfnamefont{A.}~\bibnamefont{Balogh}}, \bibinfo {author}
  {\bibfnamefont{M.}~\bibnamefont{Andre}}, \bibinfo {author}
  {\bibfnamefont{H.}~\bibnamefont{Reme}}, \bibinfo {author}
  {\bibfnamefont{M.}~\bibnamefont{Hesse}},\ and\ \bibinfo {author}
  {\bibfnamefont{C.}~\bibnamefont{Mouikis}},\ }%
  \bibfield{journal}{%
  \bibinfo {journal} {J.\ Geophys. \ Res.}\ }%
  \textbf{\bibinfo {volume} {110}},\ \bibinfo {pages} {A09206} (\bibinfo {year}
  {2005})%
  \bibAnnoteFile{NoStop}{wygant05}%
\bibitem{borg05}%
  \BibitemOpen
  \bibfield{author}{%
  \bibinfo {author} {\bibfnamefont{A.~L.}\ \bibnamefont{Borg}}, \bibinfo
  {author} {\bibfnamefont{M.}~\bibnamefont{Oieroset}}, \bibinfo {author}
  {\bibfnamefont{T.~D.}\ \bibnamefont{Phan}}, \bibinfo {author}
  {\bibfnamefont{F.~S.}\ \bibnamefont{Mozer}}, \bibinfo {author}
  {\bibfnamefont{A.}~\bibnamefont{Pedersen}},\ and\ \bibinfo {author}
  {\bibfnamefont{C.}~\bibnamefont{Mouikis}},\ }%
  \bibfield{journal}{%
  \bibinfo {journal} {Geophys.\ Res. \ Lett.}\ }%
  \textbf{\bibinfo {volume} {12}},\ \bibinfo {pages} {L19105} (\bibinfo {year}
  {2005})%
  \bibAnnoteFile{NoStop}{borg05}%
\bibitem{asano04}%
  \BibitemOpen
  \bibfield{author}{%
  \bibinfo {author} {\bibfnamefont{Y.}~\bibnamefont{Asano}}, \bibinfo {author}
  {\bibfnamefont{T.}~\bibnamefont{Mukai}}, \bibinfo {author}
  {\bibfnamefont{M.}~\bibnamefont{Hoshino}}, \bibinfo {author}
  {\bibfnamefont{Y.}~\bibnamefont{Saito}}, \bibinfo {author}
  {\bibfnamefont{H.}~\bibnamefont{Hayakawa}},\ and\ \bibinfo {author}
  {\bibfnamefont{T.}~\bibnamefont{Nagai}},\ }%
  \bibfield{journal}{%
  \bibinfo {journal} {J. \ Geophys. \ Res.}\ }%
  \textbf{\bibinfo {volume} {109}},\ \bibinfo {pages} {A02212} (\bibinfo {year}
  {2004})%
  \bibAnnoteFile{NoStop}{asano04}%
\bibitem{ren05}%
  \BibitemOpen
  \bibfield{author}{%
  \bibinfo {author} {\bibfnamefont{Y.}~\bibnamefont{Ren}}, \bibinfo {author}
  {\bibfnamefont{M.}~\bibnamefont{Yamada}}, \bibinfo {author}
  {\bibfnamefont{S.}~\bibnamefont{Gerhardt}}, \bibinfo {author}
  {\bibfnamefont{H.}~\bibnamefont{Ji}}, \bibinfo {author}
  {\bibfnamefont{R.}~\bibnamefont{Kulsrud}},\ and\ \bibinfo {author}
  {\bibfnamefont{A.}~\bibnamefont{Kuritsyn}},\ }%
  \bibfield{journal}{%
  \bibinfo {journal} {Phys.\ Rev. \ Lett.}\ }%
  \textbf{\bibinfo {volume} {95}},\ \bibinfo {pages} {055003} (\bibinfo {year}
  {2005})%
  \bibAnnoteFile{NoStop}{ren05}%
\bibitem{hesse01a}%
  \BibitemOpen
  \bibfield{author}{%
  \bibinfo {author} {\bibfnamefont{M.}~\bibnamefont{Hesse}}, \bibinfo {author}
  {\bibfnamefont{J.}~\bibnamefont{Birn}},\ and\ \bibinfo {author}
  {\bibfnamefont{M.}~\bibnamefont{Kuznetsova}},\ }%
  \bibfield{journal}{%
  \bibinfo {journal} {J.\ Geophys. \ Res.}\ }%
  \textbf{\bibinfo {volume} {106}},\ \bibinfo {pages} {3721} (\bibinfo {year}
  {2001})%
  \bibAnnoteFile{NoStop}{hesse01a}%
\bibitem{daughton06}%
  \BibitemOpen
  \bibfield{author}{%
  \bibinfo {author} {\bibfnamefont{W.}~\bibnamefont{Daughton}}, \bibinfo
  {author} {\bibfnamefont{J.}~\bibnamefont{Scudder}},\ and\ \bibinfo {author}
  {\bibfnamefont{H.}~\bibnamefont{Karimabadi}},\ }%
  \bibfield{journal}{%
  \bibinfo {journal} {Phys.\ Plasmas}\ }%
  \textbf{\bibinfo {volume} {13}},\ \bibinfo {pages} {072101} (\bibinfo {year}
  {2006})%
  \bibAnnoteFile{NoStop}{daughton06}%
\bibitem{jain09}%
  \BibitemOpen
  \bibfield{author}{%
  \bibinfo {author} {\bibfnamefont{N.}~\bibnamefont{Jain}}\ and\ \bibinfo
  {author} {\bibfnamefont{A.~S.}\ \bibnamefont{Sharma}},\ }%
  \bibfield{journal}{%
  \bibinfo {journal} {Phys.\ Plasmas}\ }%
  \textbf{\bibinfo {volume} {16}},\ \bibinfo {pages} {050704} (\bibinfo {year}
  {2009})%
  \bibAnnoteFile{NoStop}{jain09}%
\bibitem{attico00}%
  \BibitemOpen
  \bibfield{author}{%
  \bibinfo {author} {\bibfnamefont{N.}~\bibnamefont{Attico}}, \bibinfo {author}
  {\bibfnamefont{F.}~\bibnamefont{Califano}},\ and\ \bibinfo {author}
  {\bibfnamefont{F.}~\bibnamefont{Pegoraro}},\ }%
  \bibfield{journal}{%
  \bibinfo {journal} {Phys.\ Plasmas}\ }%
  \textbf{\bibinfo {volume} {7}},\ \bibinfo {pages} {2381} (\bibinfo {year}
  {2000})%
  \bibAnnoteFile{NoStop}{attico00}%
\bibitem{storey1953}%
  \BibitemOpen
  \bibfield{author}{%
  \bibinfo {author} {\bibfnamefont{L.~R.~O.}\ \bibnamefont{Storey}},\ }%
  \bibfield{journal}{%
  \bibinfo {journal} {Phil. \ Trans. \ Roy. \ Soc. \ Ser. A}\ }%
  \textbf{\bibinfo {volume} {246}},\ \bibinfo {pages} {113} (\bibinfo {year}
  {1953})%
  \bibAnnoteFile{NoStop}{storey1953}%
\bibitem{singh2011}%
  \BibitemOpen
  \bibfield{author}{%
  \bibinfo {author} {\bibfnamefont{N.}~\bibnamefont{Singh}},\ }%
  \bibfield{journal}{%
  \bibinfo {journal} {Phys. \ Rev. \ Lett.}\ }%
  \textbf{\bibinfo {volume} {107}},\ \bibinfo {pages} {245003} (\bibinfo {year}
  {2011})%
  \bibAnnoteFile{NoStop}{singh2011}%
\bibitem{singh2007}%
  \BibitemOpen
  \bibfield{author}{%
  \bibinfo {author} {\bibfnamefont{N.}~\bibnamefont{Singh}},\ }%
  \bibfield{journal}{%
  \bibinfo {journal} {J.\ Geophys. \ Res.}\ }%
  \textbf{\bibinfo {volume} {112}},\ \bibinfo {pages} {A07209} (\bibinfo {year}
  {2007})%
  \bibAnnoteFile{NoStop}{singh2007}%
\end{thebibliography}
\end{document}